\documentclass[fleqn,twoside]{article}
\usepackage[headings]{espcrc2}
\usepackage{algorithm2e}

\readRCS
$Id: espcrc2.tex,v 1.2 2004/02/24 11:22:11 spepping Exp $
\usepackage{graphicx, balance}
\usepackage[figuresright]{rotating}

\newcommand{\AmS}{{\protect\the\textfont2
  A\kern-.1667em\lower.5ex\hbox{M}\kern-.125emS}}

\title{\textbf{Forecasting Stock Time-Series using Data Approximation and Pattern Sequence Similarity}}
\author{Vishwanath R H\address[uvce]{Department of Computer Science and Engineering, 
University Visvesvaraya College of \\~Engineering, Bangalore University, Bangalore. Contact: vishwa.gld@gmail.com }, Leena S\addressmark, Srikantaiah K C\addressmark, K Shreekrishna Kumar\address{Director, All India Council for Technical Education, SWRO, Bangalore, India}, P Deepa Shenoy\addressmark[uvce], Venugopal K R\addressmark[uvce], S S Iyengar\address{Director and Ryder Professor, Florida International University, USA}  and
L M Patnaik\address{Honorary Professor, Indian Institute of Science, Bangalore 560 001, India}}
\runtitle{Prediction of Stock Time-Series Data }
\runauthor{Vishwanath R Hulipalled {\it et al.}}

\begin{document}
\begin{abstract}
  Time series analysis is the process of building a model using statistical techniques to represent characteristics of time series data.  Processing and forecasting huge time series data is a challenging task.  This paper presents Approximation and Prediction of Stock Time-series data $(APST)$, which is a two step approach to predict the direction of change of stock price indices. First,  performs data approximation by using the technique called Multilevel Segment Mean ($MSM$). In second phase, prediction is performed for the approximated data using Euclidian distance and Nearest-Neighbour technique. The computational cost of data approximation is $O(n*n_i)$ and computational cost of prediction task is $O(m * |NN|)$. Thus, the accuracy and the time required for prediction in the proposed method is comparatively efficient than the existing Label Based Forecasting ($LBF$) method \cite {R1}.
\end{abstract}

% typeset front matter (including abstract)
\maketitle

\section{INTRODUCTION}

Data mining is the process of extracting knowledge, by dredging the data from huge database. Sequence database consists of sequence of ordered events with or without notion of time. Time series data is a sequence database which consists of sequences of values or events obtained over repeated measurements of time, which can be used in prediction of any future events for user applications. Forecasting  is the prediction of forth coming events based on historical events. The recurring intervals for forecasting is based on the duration observed, $i.e.,$ it requires   many years for long term prediction, a year or more for medium term prediction and weeks or days for short term prediction.

\subsection{Motivation}
The main motivation behind this work is that, it is very much crucial for the stock market investors to estimate the behavior or trend of the stock market prices as precisely as possible in order to reach the best trading decisions for their investments. 
On the other hand, the complexity of many financial market is based on the nonlinearity and nonparametric nature of the variables influencing the index movement directions including human psychology and political events.
 The unpredictable volatile market index makes it a highly challenging task to accurately forecast its path of movement.  In this context, it is required to build an efficient forcasting model, so that the investor can utilize the most accurate time series forecasting model to maximize the profit or to minimize the risk.
\subsection{Methodologies}
 In this paper, we are using sliding window model to analyze stock time-series data. The basic idea is that rather than running computations on the entire data, we can make decisions based only on recent data. More formally, at every time $t$, a new data element arrives. This element expires at $(t + w)$, where $w$ is the window size or length. The sliding window model is useful for moving object search, stock analysis or sensor network analysis, where only recent events may be important and reduces memory requirements because only a small window of data is used.

\subsection{Contribution} 
%This work is based on the concept used in the LBF algorithm discussed in \cite {R1}, which uses only  labels associated with each pattern, without using the real values until the last step of the prediction task using clustering method.
 In this paper, a new method called $APST$ has been proposed, that generates the predicted values for the original stock time series data. Here, we first perform preprocessing upon the historical stock time series data to generate the sequence of approximated values using Multi scale Segment Mean approach \cite {R2}. Then, we use these approximated sequence of values for the predicting process. To forecast, we use the Euclidian distance approach to find the nearest neighbor objects to identify the  similar set of objects as used in \cite {R3}. The accuracy of $APST$ is estimated by computing the percentage of  error based on the difference between the predicted value and the actual known value for each test samples.
\subsection{Organization}
The rest of the paper is organized as follows, Section 2  discusses briefly the Literature on  stock  price time series forecasting. Section 3 presents the background work, Section 4 contains Problem definition, Section 5 describes the  System Architecture, section 6 presents the Mathematical model and  Algorithm, Section 7 addresses the Experimental Results for proposed  method and existing $LBF$ technique. Concluding remarks are summarized in the Conclusion.
\section{LITERATURE SURVEY}
%In most of the survey, several techniques have been proposed on forecasting. Time series forecasting is the most common and has the wide number of applications in various fields. 

%\vskip 2mm 

% Machine learning is a well known algorithm, which is used in wide range of applications that has been extensively studied for its potentials in prediction of financial markets.
 Popular algorithms like Support Vector Machine ($SVM$) and Reinforcement learning, are effective in tracing the stock market and helps in maximizing the profit of stock option purchase while keeping the risk low \cite{R4}-\cite{R5}.
%\vskip 2mm
Nayak \textit et al., \cite{R6} tested  the predictive power of the clustering technique on Australian stock market data using a brute force method. This is based on the idea that a cluster formed around an event could be used as a good predictor for the future event. 
%\vskip 2mm
%Reinald et al., in \cite{Ref4} proposed an approach to predict next-day electricity prices based on the Generalized AutoRegressive Conditional Heteroskedastic (GARCH) methodology that is used to analyze time series data. This paper focuses on day-ahead forecasts of electricity prices with high volatility periods using a GARCH methodology approach. Based on the historical data value GARCH models provide 24-hour forecasts of the next day.
\vskip 2mm
 Conejo \textit et al., \cite{R7}  proposed a technique to forecast day-ahead electricity prices based on the wavelet transform and $ARIMA$ models.
The series of prices is decomposed using the wavelet transform into a set of constitutive series. Then, the $ARIMA$ models are used to forecast the future values of this consecutive series. In turn, through the  inverse wavelet transform, the $ARIMA$ model reconstructs the future behavior of the price series and therefore to forecast prices.
%vskip 2mm
Akinwale \textit et al., in \cite {R8} used NN approach to predict the untranslated and translated Nigeria Stock Market Price ($NSMP$). They used 5-$j$-1 network topology to adopt the five input variables. The number of hidden neurons determined the $j$ variables during the network selection. Both the untranslated and translated statements were analyzed and compared. The performance of translated $NSMP$ using regression analysis or error propagation was more superior to untranslated $NSMP$. The result was showed on untranslated $NSMP$ ranged for 11.3\% while 2.7\% for $NSMP$.

\vskip 2mm
Kuang \textit et al., \cite{R9} used the $MARX$ (Moving average AutoRegressive eXogenous  prediction model) fusion with $RS$ (Rough Set theory) and $GS$ (Grey System theory)  to create an automatic stock market forecasting and portfolio selection mechanism.  Financial data were collected automatically every quarter and are input to an $MARX$ prediction model for forecasting the future trends. Clustered using a K means clustering algorithm and then supplied to a RS classification module which selects appropriate investment stocks by a decision-making rules.
 The advantages are combining different forecasting techniques to improve the efficiency and accuracy of automatic prediction. Efficacies of the combined models are evaluated by comparing the forecasting accuracy of the $MARX$ model with $GM$ (1, 1) model. The hybrid model provides a high accuracy.

%\vskip 2mm
%Abdulsalam Sulaiman Olaniyi \textit et al., \cite{R10} used the moving average [MA] method to uncover the patterns, relationship and to extract values of variables from the database to predict the future values of other variables through the use of time series data. The advantage of the MA method is a device for reducing fluctuations and obtaining trends with a fair degree of accuracy. %This techniques proven numeric forecasting method using regression analysis with the input of financial information obtained from the daily activity equities published by Nigerian stock exchange. 

\vskip 2mm
Suresh \textit et al., \cite{R10} used the data mining techniques to uncover the hidden pattern, predict future trends and behaviors in financial market.
Martinez \textit et al.,  \cite{R11} proposed the nearest neighbor technique called Pattern Sequence-based Forecasting ($PSF$). This method uses clustering technique to generate labels and makes predictions basing only on these labels. However, it is quite difficult to determine the suitable number of clusters in the clustering step and in some anamoly cases, if samples are not in the training set. This method cannot predict events in the future even when the length of a label pattern is one. The proposed work is the extension of our own work discussed in \cite{R12}. 
% In this paper, author applied ant algorithm to accommodate a flexible and dynamic pattern-matching task in time series analysis. Apart from segment size the ant to sub-time-series size affects the system performance. %In this paper, the ratio was set to 1 and also the ratio reduced to obtain a better result.

%\vskip 2mm
% This approach gives a new methodology to automate the obtaining of the labels which supplies rules for forecasting.% And the approach was applied to forecast electricity market prices and electricity demand \cite{R13}. 

\section{BACKGROUND}  
%F. Martınez-Alvarez \textit et al., [1] proposed an LBF algorithm for predicting the electricity price of belongs to three major markets, the Spanish electricity market operator (OMEL), New York Independent system operator (NYISO), and the Australia’s National Electricity Market (ANEM).
The Label Based Forecasting ($LBF$) \cite{R1} algorithm consists of two phases. In the first phase, a clustering technique is used to generate the labels and in the second phase, forecasting is performed by using the information provided by clustering. In $LBF$ method, it is quite difficult to determine the suitable number of clusters in clustering step and they are not using the actual values of the data set for the prediction, instead they use the set of labels created by clustering approach, this may lead to errors in prediction.
\vskip 2mm
In the proposed method, we use real values of the input time series instead of labels for the prediction process. $APST$ algorithm first performs the data approximation by using the technique called Multilevel Segment Mean ($MSM$) and in the second phase, prediction is performed for the approximated data.

\section{PROBLEM DEFINITION}
\subsection{Problem Statement}

Let, $ $P$($i$) \in V^{d}$  be a vector composed of the  daily closing stock prices of a particular company, corresponding to $d$ number of days, which is given by the equation,
\begin{equation} 
  P(i)= [p_1, p_2, \ldots, p_{d}]  
\end{equation} 
then, approximate the vector content $V^{d}$ to get the approximated stream of data, $i.e.,$
\begin{equation} 
  Ap= [Ap_1, Ap_2, \ldots, Ap_n] 
\end{equation}
The objective is to predict the $(d+1)^{th}$  day stock  price by searching the $K$ nearest neighbour in $Ap$. 
\subsection{Assumptions}
i) We divide $N$ days stock prices into equal number of $K$ groups. In our example, we consider size of each $K$ groups to be 27 consecutive elements from the input data stream $D$.\\
 ii) We further divide each $K$ groups into $t$ number of subsegments. In our example, we consider size of each $t$ subsegments to be 3 consecutive  elements from each $K$ groups.\\

%\subsection{Objectives}
The obective is to forecast the Stock time series data by finding  similar patterns over a stream of stock time series data and reduce the processing cost and dimensionality of time series $W_i$ and pattern $p_j$.

\section{SYSTEM ARCHITECTURE}
The system architecture consists of the following components, (i) Data source, (ii) Data Approximation Process, (iii) Prediction Process and  (iv) Predicted  Data Set. The complete architecture is as shown in the Figure 1.\\

\textbf{\textit{Data Source:}} It is the collection of historical stock price time series data. In this the closing  stock price values of many companies  for several years are  collected and are stored in historical data base.\\

\textbf{\textit{Data Approximation Process:}} This is a preprocessing step for the prediction task. In this, the original input stock time series data is approximated using the $MSM$ technique which is discussed in detail further and an example is shown in Figure 2 and the data approximation steps are discussed in detail, in $Phase-1$ of the algorithm $APST$, as shown in the Table 1. The main objective of this step is to condense the data set. \\

\textbf{\textit{Prediction Process:}} The main goal of this paper is to forecast the stock time series data. It involves the three steps: i)Finding $K$ Nearest Neighbours, ii)Selecting $K$-elements following each Nearest Neighbours, lastly iii)Finding the mean of the $K$-elements. Figure 3 shows the example for the prediction process and the detail steps of the prediction process is  discussed in $Phase-2$ of the algorithm $APST$, as shown in the Table 1.\\

\textbf{\textit{Predicted Data Set:}} This is the output obtained from the prediction process and is the collection of the predicted values which are later compared with the origional stock time series values to evaluate the prediction accuracy of the proposed model.

\vskip 2mm
\begin{figure*}[ht!]
\begin{center} 
\includegraphics[width=10.5cm,height=7cm]{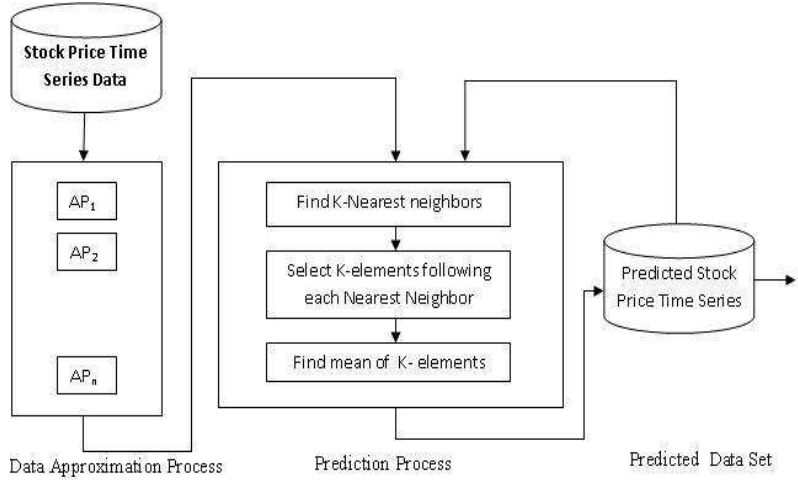}
\caption{System Architecture}
\end{center}
\end{figure*}

\section{MATHEMATICAL MODEL AND ALGORITHM}
\subsection{Data Approximation Process} 
The given stock time series data $D$ of size $N$ is divided into number of equal partitions, $P_1$, $P_2$, \ldots, $P_{n}$ and the total number of partitions is given by
\begin{equation}
  n= N/K 
\end{equation} where, $K$ is size of each partition.
For each partition $P_i$, where $ 1 \leq  $ i $ \leq $ K, $ $ segment $P_i$ into $n_i$ segments, $S_1$,\ldots,$S_{n_i}$ and the total number of segments is given by,
\begin{equation}
% $n_i$= $|P_k|/$$t$,
 n_i= |P_k|/t,
\end{equation}
 where, $t$ is size of each segment.
The set of the segments for each partition $P_i$ is given by
\begin{equation}
S_{p_i}=[S_1, S_2,..., S_{n_i}]
\end{equation}
The data approximation for the input stock time series data $D$ is obtained by computing the segments mean from $level$ $l$ to $level$ $0$ in the tree
 and the total number of levels in the tree is computed by following equation,
\begin{equation}
 n_l=\log_t K
\end{equation} 
and at each $level$ $j$ we can form $(3^l)$ disjoint segments, for each segment $ S_j \in S_{p_i} $  in the $level$ $ $l$= \log_t $K$ $, mean of all the elements in $S_i$ is computed and stored in $AP_{ij}$[$l-1$]. The mean value in the [$l-1$] levels are grouped as one segment, again considering each segment in [$l-1$] level, find the mean of all elements in that segment and stored in $AP_{ij}$[$l-2$]. Similar procedure is followed to obtain $AP_{ij}$[$l-3$] upto $AP_{ij}$[$l-l$] $i.e.,$ $AP_{ij}$[$0$]. $AP_{ij}$[$0$] gives the approximated value of level $j$ for the partition $P_i$.  This process is continued for computing the approximation for all the levels, finally, the approximated values are as follows,\\
 $Ap$= [$Ap_{10}$, $Ap_{20}$, \ldots, $Ap_{n0}$].\\
Figure 1 shows the segment mean representation of stock time series data $D$ of length $N$.\\

\begin{figure*}[ht!]
\begin{center} 
\includegraphics[width=10cm,height=5cm]{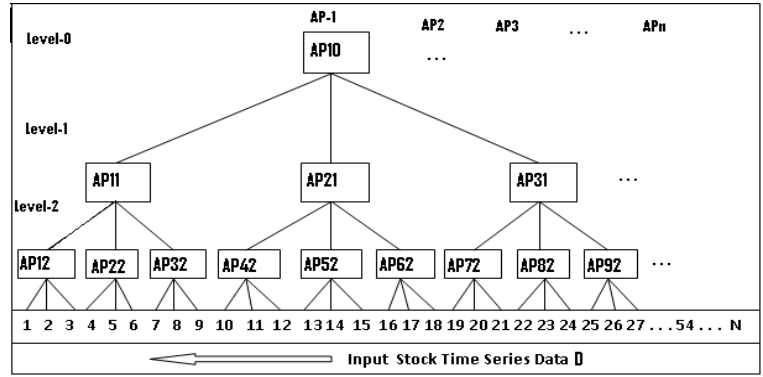}
\caption{Data Approximation using $MSM$}
\end{center}
\end{figure*} 
\textit{Example:}
  This example shows the computation of the approximation values for the given input stock time series data. In Figure 2, at  $level$ 2, 1 and 0 we construct total of 9, 3 and 1 segments respectively.
The first segment $AP_{12}$ at $level$ 2 is the mean of the $1^{st}$, $2^{nd}$ and $3^{rd}$ value in the input stock time series data $D$. Similarly, the second segment $AP_{22}$ at $level$ 2 is the mean of the $4^{th}$, $5^{th}$ and $6^{th}$ values in the input stock time series data $D$.  So we can construct the segments  upto $AP_{92}$ on $level 2$. $i.e.,$  $AP_{12}$, $AP_{22}$,  $AP_{32}$ \ldots, $AP_{92}$ as shown in Figure 2.\\
At $level$ 1, we construct 3 segments $AP_{11}$, $AP_{21}$ and $AP_{31}$ as follows:\\
The segment $AP_{11}$ is computed by the mean of 3 adjacent segments on $level$ 2, $i.e.$, 
$AP_{11}$=[$AP_{12}$ + $AP_{22}$ + $AP_{32}$]/3, $AP_{21}$=[$AP_{42}$ + $AP_{52}$ + $AP_{62}$]/3 and $AP_{31}$=[$AP_{72}$ + $AP_{82}$ + $AP_{92}$]/3.
\vskip 2mm
 Lastly, we can compute the only one segment $AP_{10}$  at $level$ $0$ as follows,\\
$AP_{10}$=[$AP_{11}$ + $AP_{21}$ + $AP_{31}$]/3. Hence $AP_{10}$ is the approximated value of the first group $K_1$ in the input stock time series data $D$, $i.e.$, $AP_{1}$=$AP_{10}$.
In the same manner, we can compute another approximated value $AP2$ from the next partition $K_2$ and next approximatd value $AP3$ from the next partition $K_3$ and so on. The set of approximated values at $level$ $0$, are formed as follows, $Ap$= [$Ap_{10}$, $Ap_{20}$, \ldots, $Ap_{n0}$]

\subsection{Prediction Process}
The Stock price time series values are predicted using the data approximation values obtained in the previous section. 

%\vskip 2mm
\begin{figure*}[ht!]
\begin{center} 
\includegraphics[width=10.5cm,height=6.75cm]{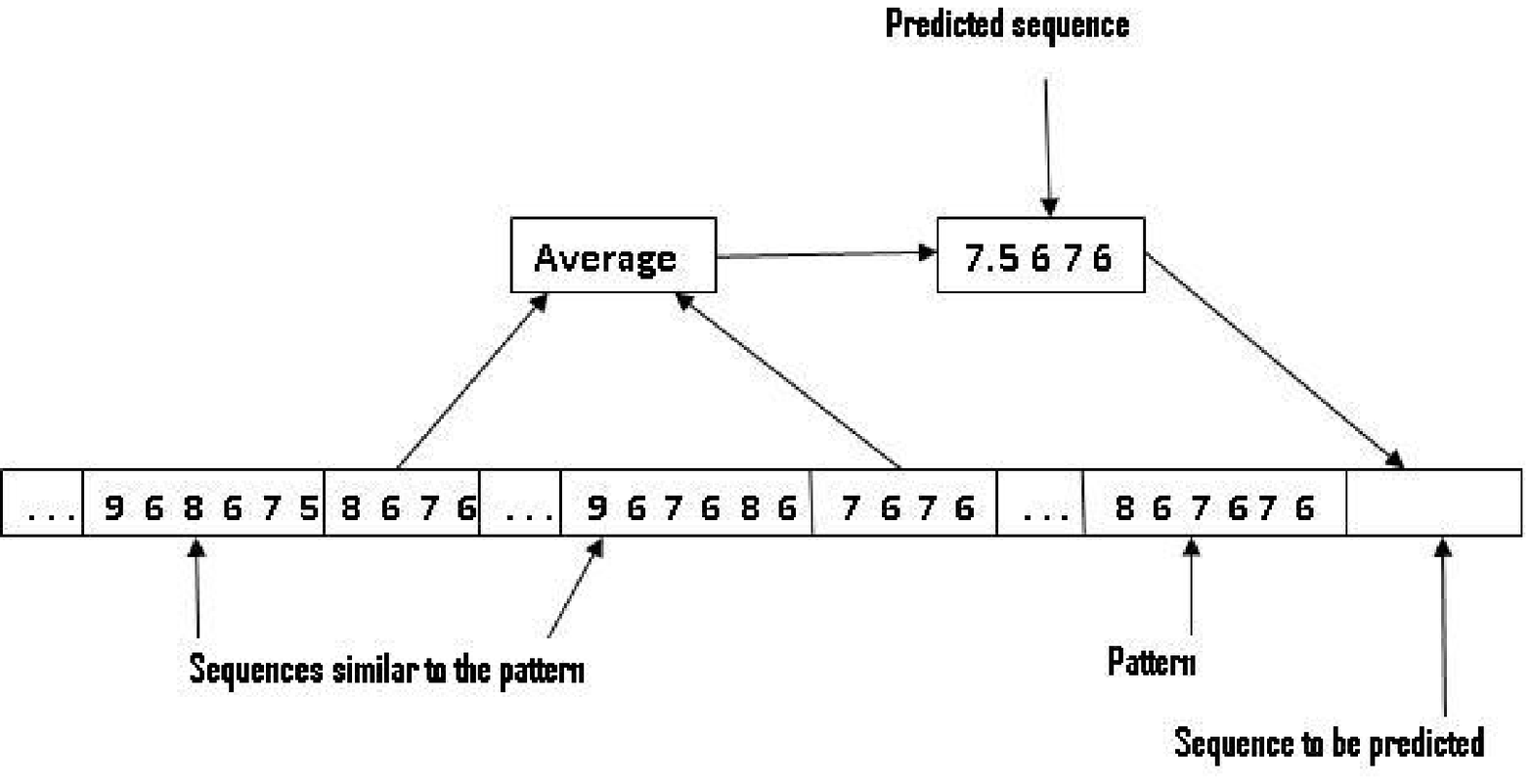}
\caption{Prediction Process using Nearest Neighbour}
\end{center}
\end{figure*}

%\textbf{\textit{Predicted Data set:}}
%Finally we get the predicted stock price for the Original input stock stream time series data after the end of the prediction process. These predicted values and Original values are further considered for the comparative analysis.\\

%\section {Mathematical Model}
%\subsection{Data Approximation}

%Let $D$ be the data set, and size of the $D$ is , $N$= $|D|$. The number of partitions is given by $n= N/K$ 
%where $K$ is size of each partition. Now let us partition $D$ into $P_1$,$P_2$, . . .$P_n$. To do data aggregation we can perform following steps:\\
%For each partition $P_i$, $1 \leq  $i$ \leq  $n$ $, Now, Segment each partition $P_i$, into $n_i$ segments.\\ where $n_i$=$|P_i|$/$t$ , $t$ is the size of each segment.\begin{equation}  %$Sp$_i$= [$S_1$, $S_2$, \ldots, $Sn_i] Sp_i= [S_1, S_2, \ldots, Sn_i] \end{equation}  
 %For number of levels, in the tree is given by
%\begin{equation} 
%$ l=\log_t K $
%\end{equation}  
%For each level
%\begin{equation} 
%l=\log_t K ; k>0 ; k--
%\end{equation}  
%For each Segment
%\begin{equation} 
%S_j \in S[P_i] ,   1\leq j \leq n_i
% 1\leqslant j\leqslant n_i
%\end{equation}  

%Now find the mean of elements in the segment $S_j$ and store in the vector $Ap_j$[$l$-1].
%Similarly, finding $Ap_j$[$l$-2],$Ap_j$[$l$-3] \ldots $Ap_j$[0]=$Av_p$ 
%Finally, the approximated values in segment $S$[$P_i$] is given by 
%\begin{equation}
%Ap={Av_p1,Av_p2, \ldots, Av_pn}
%\end{equation}

%\subsection{Prediction Process}
\vskip 2mm
Given the set of approximated values as
%\begin{equation}
$ Ap={Ap_1,Ap_2, \ldots, Ap_n}$
%\end{equation}
and we need to compute a set of predicted values as follows,
\begin{equation}
P'={v_1,v_2, \ldots, v_m}
\end{equation}
 Let $W$ be the window of size $w$, and $m$ be the size of the predicted values, in our case $m$=1.
Now consider the last $w$ elements in the input sequence $Ap$, $i.e.$, pattern set $PS$ and is given by,
\begin{equation}
PS={AP_{n-w},AP_{n-w-1}, \ldots, AP_{n-1},AP_n}
\end{equation}
Next, the nearest neighbour in $AP$ is obtained for $PS$. Let 
 $k$ be number of nearest neighbours in $AP$. In $PS$, the set of nearest neighbour is given by,
\begin{equation}
NN={nn_1,nn_2, \ldots, nn_k}
\end{equation}

For each nearest neighbour, $nn_i \in NN $, retrive sequence of $m$ elements next to $nn_i$  $i.e,$.
\begin{equation}
EL_i={e_{i1},e_{i2}, \ldots, e_{im}}
\end{equation}
are the sequence of elements next to $nn_i$.
Set of sequence of $m$ elements next to all the nearest neighbours in $NN$ is given by the set $NS$. 
\begin{equation}
%NS={a_{i1},a_{i2}, \ldots, a_{im}} , 1\leq i \leq k
NS=EL_1,EL_2, \ldots EL_k
\end{equation}
The predicted value in the sequence, that consists of average of corresponding elements in the set $NS$ is given by,\\
$P'<{v_1,v_2, \ldots, v_m}> =$
\begin{equation}
{\frac 1 k{\sum_{i=1}^k E_{i1}},\frac 1 k{\sum_{i=1}^k E_{i2}},  \ldots, \frac 1 k{\sum_{i=1}^k E_{im}}}
\end{equation}
\textit{Example:} Prediction process is shown in the Figure 3. Table 1 shows the complete algorithm for Data approximation and Prediction of Stock time series data. 
\begin{table*}
\centering
\caption{Algorithm :$APST$: Approximation and Prediction of Stock Time-Series Data}
\begin{tabular}{|p{13cm}|}
\hline
$\mathbf{Algorithm:}$ $\mathbf{\it APST(D,~K,~,t~,w,~m)}$\\
\textbf{Input:}
\indent\hspace{0.0cm}	 	 $D$	:	 Stock time series data set of size $N$ \\
\indent\hspace{1.5cm}            $k$	: 	  Size of each partition\\
\indent\hspace{1.5cm}            $t$	:     Size of each segment\\
\indent\hspace{1.5cm} 		 $w$	:    Window size  \\
\indent\hspace{1.5cm}            $m$	:	 Size of sequence to predicted\\
 \textbf{Output:} 
\indent\hspace{0.0cm}  		$P'$	:	 Predicted Stock values 	\\ \\
\textbf{----------------------------------------------------------------------------------------------}\\
 \textbf{Phase-1:} $Data$  $Approximation$\\
\textbf{----------------------------------------------------------------------------------------------}\\
\textbf{begin}\\
\indent\hspace{0.5cm} $N=1$, $n=N/K$, $l=\log_t k$;\\
\indent\hspace{0.5cm} $P$= Partition $D$ into $p_1$, $p_2$, \ldots, $p_n$ of size $k$\\
 \indent\hspace{0.5cm}\textbf{for each} partition $p_i$ $\in$ $P$ \textbf{do}\\
 \indent\hspace{0.9cm}Segment $p_i$ into $n_i$ segments $s_1$, $s_2$, \ldots, $s_{ni}$\\
 \indent\hspace{0.9cm} $S_{P_i}$={$s_1$, $s_2$, \ldots, $s_n$}\\
 \indent\hspace{1.2cm}\textbf{for each} Segment $S_j$ $\in$ $S_{P_i}$ \textbf{do} \\
\indent\hspace{1.5cm}  $AP_{ij}[l-1]$=Mean of each elements in $S_j$ \\
 \indent\hspace{1.2cm}\textbf{end for}\\
\indent\hspace{0.5cm}\textbf{end for}\\
 \indent\hspace{1.5cm}\textbf{for} $l=(\log_t k,l\geq 0, l-- )$ \textbf{do}\\
%$l$=$(\log_t k,l \geqslant 0,l--)$ \textbf{do}\\
%\indent\hspace{1.5cm}\textbf{for} $l=(\log_t k$,l $\geqslant$ 0,$l--$) \textbf{do}\\ 
 \indent\hspace{1.8cm}Groups the elements in $AP_{ij}[l-1]$ into segments of $t$ elements\\
\indent\hspace{1.8cm} for each segment find the mean and store in $AP_{ij}[l-2]$\\
 \indent\hspace{1.8cm}repeating the same steps to find the mean upto $AP_{ij}[l-l]$\\
\indent\hspace{1.5cm}\textbf{end for}\\ 
\textbf{end}\\ \\
 //The final set of approximation values for all the levels are,\\
//$AP_{ij}[l-l]$ = $AP_{ij}[0]$ = $Ap$= [$Ap_{10}$, $Ap_{20}$, \ldots, $Ap_{n0}$] \\
//These values are used in the Phase-2 for the Prediction.\\
 \textbf{---------------------------------------------------------------------------------------------}\\
 \textbf{Phase-2}:  $Data$  $Prediction$\\
\textbf{----------------------------------------------------------------------------------------------}\\
\textbf{begin}\\
$Ap$= [$Ap_{10}$, $Ap_{20}$, \ldots, $Ap_{n0}$]\\
 \indent\hspace{0.5cm}$PS=<{AP_{n-w},AP_{n-w-1}, \ldots, AP_{n-1},AP_n}>$ \\
\indent\hspace{0.5cm}$NN$= Find the nearest neighbours for $PS$ in $AP$  $<{ nn_1, nn_2, \ldots, nn_k}>$\\
\indent\hspace{0.5cm} \textbf{for each} $nn_i$ $\in$ $NN$ \textbf{do}\\
\indent\hspace{0.9cm} $E_i$= Extract Sequence  $<{ e_{i1}, e_{i2}, \ldots, e_{im}}>$ of $m$ elements next to $nn_i$ \\
\indent\hspace{0.5cm} \textbf{end for}\\
\indent\hspace{0.5cm} \textbf{for each} $j$=1 to $m$ \textbf{do}\\
 \indent\hspace{1.0cm}\textbf{for each} Element $E_i$ $\in$ $E$ \textbf{do} \\
 \indent\hspace{2.cm} $P'[j]$=$P'[j]$+$e_{ij}$\\
\indent\hspace{1.0cm}\textbf{end for}\\
\indent\hspace{0.5cm} \textbf{end for}\\
%   \textbf{end}
%\textbf{-------------------------------------------------------------------------------------------}
${\bf}$end\\
\hline
\end{tabular}
\label{tab:alg1}
\end{table*}
\vskip 2mm
The aproximated values are extracted from the stock time series data $D$. Considering the  patterns of length $w$ in the aproximated values, we have to predict a stock sequence of the next time step. In the prediction process, search for $k$
nearest neighbors stock values within the threshold $\psi$ of
that pattern and then the stock sequences next to the found
neighbors are extracted. The predicted 
stock sequence is then estimated by taking the mean of the sequences
found in the  previous step.
\vskip 2mm
The computational cost of our proposed $APST$ method, for data approximation is $O(n*n_i)$. Where, $n$ is the number of partitions and $n_i$ is the number of segments. The computational cost of prediction task is $O(m * |NN|)$. Where, $m$ is the size of sequence to be predicted and  $|NN|$ is the total size of the nearest neighbours. Thus, the accuracy and the time required for prediction in the proposed method is comparatively efficient than the existing Label Based Forecasting ($LBF$) method.
\section{EXPERIMENTAL RESULTS}
Experiments are conducted on two  real datasets,
TAIiwan stock EXchange index dataset (TAIEX) and Bombay Stock EXchange index dataset (BSEX) for different companies. The performance of our
prediction approach is compared with that of $LBF$ method. We use Mean Error Relative ($MER$) and Mean Absolute Error ($MAE$) for evaluation which are defined as follows [1].
%To assess the performance of the $APST$ approach following two quality measures are used:
%\textit{Mean Error Relative to x (MER): }
\begin{equation}
Mean Error Relative=100 {\frac 1 N} \sum_{d=1}^N \frac{\vert {P'} - {P} \vert} {\bar{P}}
\end{equation}
Where,  $P'$  is the predicted stock prices  at particular day ‘$d$’.   $P$ is the current stock prices for particular day ‘$d$’.
%\[\hat{Pr}_h\]
%\begin{equation}
$\bar{P}$ is the mean stock prices for the period of interest(day/week).    $N$ is the number of predicted days 

%\textit{Mean Absolute Error (MAE):}
\begin{equation}
Mean Absolute Error= {\frac 1 N} \sum_{d=1}^N \vert {P'} - {P} \vert
\end{equation}

\begin{table*}[ht!]
\begin{center}
\caption{The Prediction Errors by $LBF$ and $APST$ methods on the TAIEX dataset for the financial year 2010 }
\centering
\label{table1}
\begin{tabular} {|c|c|c|c|c|}
\hline
%\multicolumn{5} {c} {} \\
\cline{1-5}
Month	& MER($LBF$) & MER($APST$) 	& MAE($LBF$) & MAE($APST$)	\\
\hline
April	& 5.02	& 4.22	& 0.53  & 0.43	     \\
\hline
May	& 8.30	& 7.22  & 0.56  & 0.46		     \\
\hline
June     & 6.89	& 5.59  & 0.51	& 0.41	 \\  
\hline
July	& 7.41	& 6.21	& 0.47	& 0.37	  \\
\hline
Aug	& 8.37	& 7.57	& 0.47	& 0.37		\\
\hline
Sep	& 7.30	& 6.40	& 0.45	& 0.35		\\
\hline
Oct	& 4.62 	& 3.68	& 0.47	& 0.37	\\
\hline
Nov	& 7.26	& 6.28	& 0.44	& 0.34	\\
\hline
Dec	& 6.88	& 5.88	& 0.43	& 0.35	\\
\hline
Jan	& 7.20	& 8.26	& 0.44	& 0.36	\\
\hline
Feb	& 6.26	& 4.26	& 0.44	& 0.35	\\	
\hline
Mar	& 7.26	& 5.26	& 0.44	& 0.38	\\
\hline

%&Average MER=6.89, &Average MAE=0.47 \\
%\hline

\end{tabular}
\end{center}
\end{table*}

Using the above two equations, we compute $MER$ and $MAE$ for both the existing $LBF$ method and proposed $APST$ method for the TAIEX dataset and shown in the Table 2 and Table 3. From Table 2 and 3, it shows that the average MER is 6.89\%, Average MAE is 0.47\% in the existing $LBF$ method, whereas in the proposed $APST$ method, the average $MER$ is 5.90\% and Average $MAE$ is 0.37\% . 
\vskip 2mm
The proposed $APST$ method is $\approx$ 1\% more efficient with respect to $MER$ and 0.1 \% more efficient for $MAE$ compared to existing $LBF$ method. 

\begin{table}[ht!]
\begin{center}
\caption{Performance of $LBF$ and $APST$ with Respect to $MER$ and $MAE$}
\label{table1}
\begin{tabular} {|c|c|c|}
\hline
%\multicolumn{5} {c} {} \\
\cline{1-3}
	& AVG.MER 	 & AVG.MAE 	  \\
\hline
$LBF$	& 6.89	& 0.47	\\
\hline
$APST$	& 5.90	& 0.37   \\

\hline
%&Average MER=5.90, &Average MAE=0.37 \\
%\hline

\end{tabular}
\end{center}
\end{table}

%This section discusses about comparison betweenprediction by existing $LBF$ method and the proposed APST method on given stock time series data. 
\vskip 2mm
The graphs shown in Figures 4 and 5, indicates that, the prediction accuracy of our proposed $APST$ method is better than that of the existing $LBF$. The graph is  plotted by taking the actual stock price values against the predicted stock price values for both the methods.
\vskip 2mm
The graph shown in Figure 6, indicates that, the average CPU time required to forecast diferent stock timeseries data. We observed that, the average CPU time required for existing $LBF$ method is 0.61 miliseconds. Whereas in our proposed $APST$ method, the average CPU time required is 0.5 miliseconds. Our technique $APST$ is 0.11\% more efficient than the existing $LBF$ method because, in $LBF$ they consider the entire data set $N=|D|$, whereas in $APST$ we consider the approximated data set, $n=|AP|$ ${i.e.,}$ Size of data in $APST$=$(N/n)$. Time complexity of $LBF$ is $O(No.of days * |E_{s_d}|)$, whereas in $APST$ time complexity is $O(m*|NN|)$, and number of subsequences in $APST$ is less than the $E_{s_d}$ in $LBF$.

\begin{figure}[ht!]
\begin{center} 
\includegraphics[width=7.5cm,height=5cm]{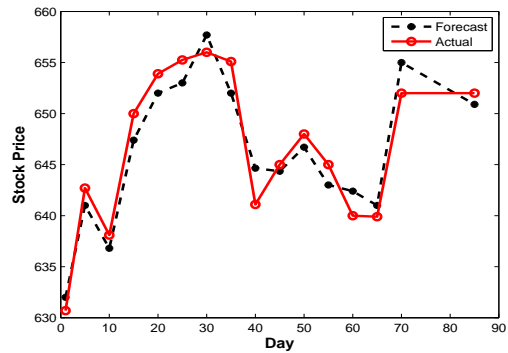}
\caption{Comparison between Actual value and Forecasted value in $LBF$
method
 }
\end{center}
\end{figure}

\begin{figure}[ht!]
\begin{center} 
\includegraphics[width=7.5cm,height=5cm]{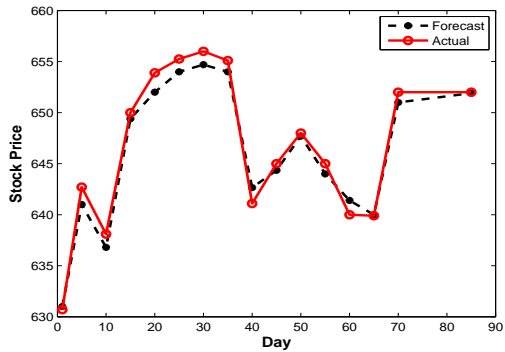}
\caption{Comparison between Actual value and Forecasted value in $APST$ method
 }
\end{center}
\end{figure}

\begin{figure}[ht!]
\begin{center} 
\includegraphics[width=7.5cm,height=5cm]{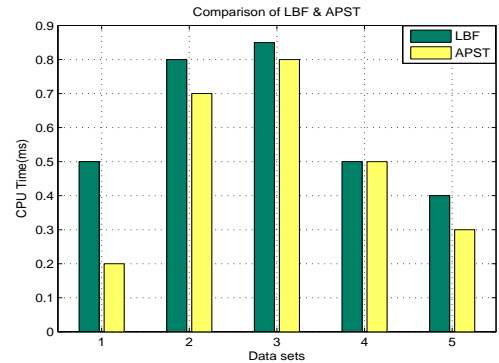}
\caption{Comparison of Forecasting Time between $LBF$ and $APST$ methods
 }
\end{center}
\end{figure}

\section{CONCLUSIONS}

The proposed mechanism $APST$ works in two phase process. In the first phase we perform data approximation using Multiscale Segment Mean($MSM$) approach to get the approximated values of the given stock time series data. In the second phase, the prediction of stock time series is carried out using the Euclidian distance and the Nearest Neighbour approach. The computational cost of proposed method with respect to data approximation is $O(n*n_i)$ and for the prediction task is $O(m * |NN|)$ respectively. 
\vskip 2mm
Further, our experimental results show that the average MER is 6.89\%, average MAE is 0.47\% in the existing $LBF$ method, whereas in the proposed method, the average $MER$ is 5.90\% and average $MAE$ is 0.37\%. Thus, the proposed method is $\approx$ 1\% more efficient with respect to $MER$ and 0.1 \% more efficient for $MAE$ compared to existing $LBF$ method.
\vskip 2mm
 Also, the average CPU time required for existing $LBF$ method is 0.61 miliseconds, whereas in the proposed method, it is 0.5 miliseconds. Thus,  proposed method is  0.11\% more efficient than the existing method. Future enhancement can be  focused on selecting the window size dynamically and fine tune the matching sequence.

\small
\balance

\noindent{\includegraphics[width=1in,height=1.8in,clip,keepaspectratio]{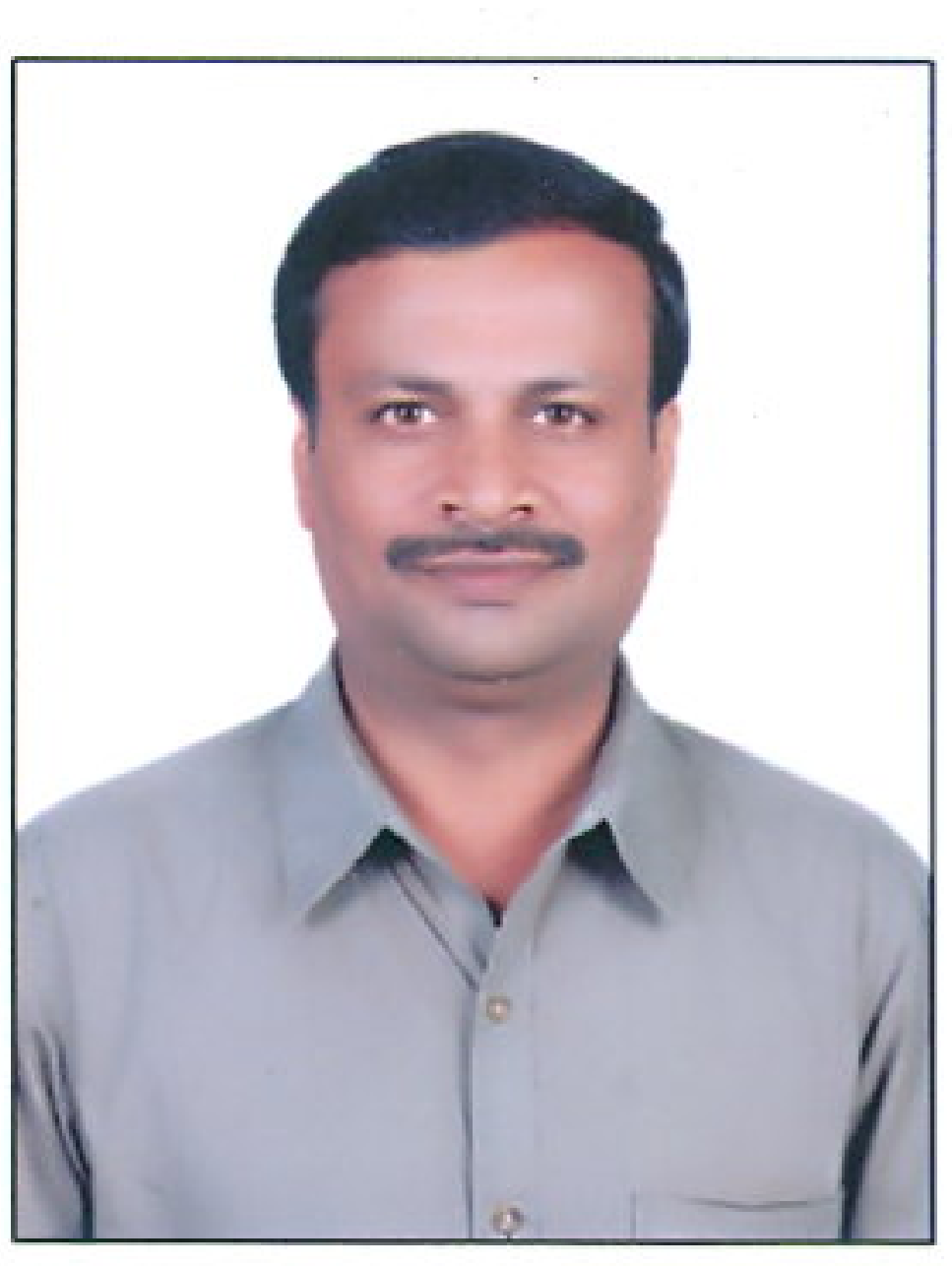}}
\begin{minipage}[b][1in][c]{1.8in}
{\centering{\bf{Vishwanath R Hulipalled}}  is an Assistant Professor in the Department of Computer Science and Engineering at Sambhram Institute of Technology, Bangalore, India. He received his Bachelors degree in Computer Science and Engineer} \\
\end{minipage}
ing from  Karnataka University and Master of Engineering from UVCE, Bangalore University, Bangalore. He is presently pursuing his Ph.D in the area of Data Mining  in JNTU Hyderabad. His research interest includes Time Series Mining and Data Analysis.\\ \\

\noindent{\includegraphics[width=1in,height=1.8in,clip,keepaspectratio]{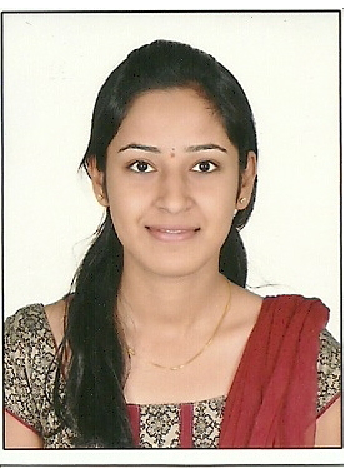}}
\begin{minipage}[b][1in][c]{1.6in}
{\centering{\bf{Leena}} is pursuing B.E in Department of Computer Science and Engineering, University Visveswaraya College of Engineering, Bangalore.} 
Her research interest is in the area of Data Mining and Time Series Mining.  \\\\\\
\end{minipage}\\ \\

\noindent{\includegraphics[width=1in,height=1.5in,clip,keepaspectratio]{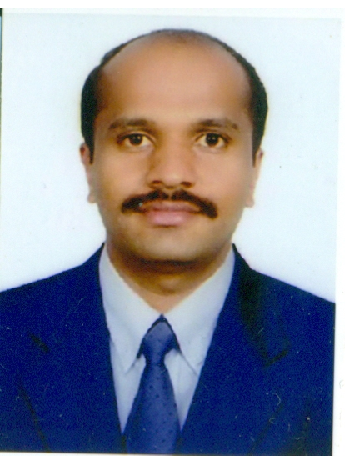}}
\begin{minipage}[b][1in][c]{1.8in}
{\centering{\bf {Srikantaiah K C }} is an Associate Professor in the Department of Computer Science and Engineering at S J B Institute of Technology, Bangalore, India. He obtained his B.E and M.E degrees in Computer Science and Engineering from Bangalore University, Bangalor-}  \\ \\ 
\end{minipage}
e.  He is presently pursuing his Ph.D programme in the area of Web Mining in Bangalore University. His research interest is in the area of Data Mining, Web Mining
and Semantic Web. \\ \\

\noindent{\includegraphics[width=1in,height=1.5in,clip,keepaspectratio]{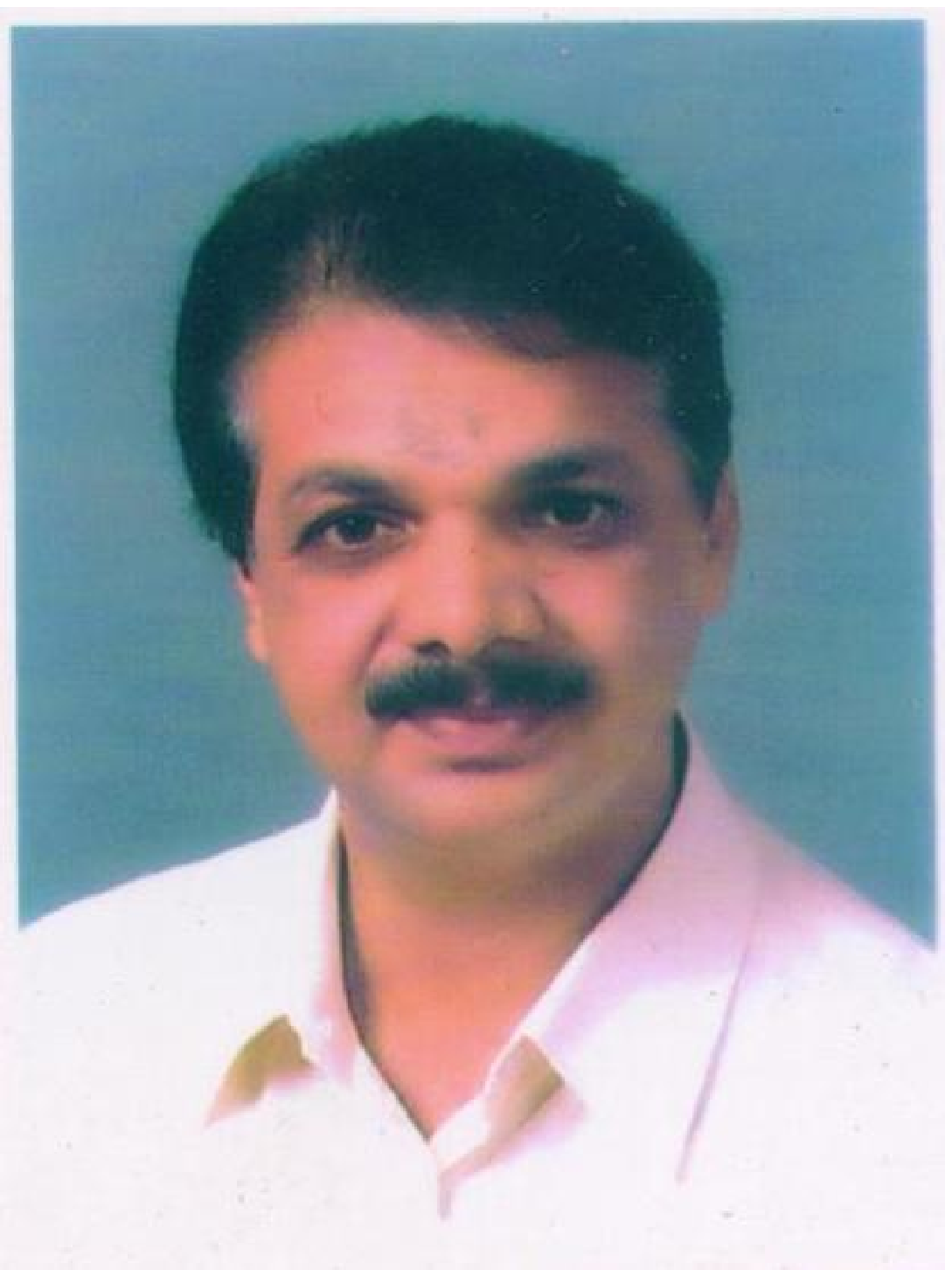}}
\begin{minipage}[b][1in][c]{1.8in}
{\centering{\bf {K Shreekrishna Kumar}} is currently the Director of All India Council for Technical Education, SWRO, Bangalore. He obtained his Master of Science from Bhopal University. He received his Masters degree in Information Technology from Punjab University. He was awa}  \\ \\ 
\end{minipage}
 rded Ph.D in Physics (Glass Technology) from Mahatma Gandhi University. He was the member of the Jury Panel, Indian Journal of Pure and Applied Physics, CSIR (New Delhi), He was the Collaborative researcher, Nuclear Science Centre, New Delhi.  \\

%\noindent{\includegraphics[width=1in,height=1.8in,clip,keepaspectratio]{sk.eps}}
%\begin{minipage}[b][1in][c]{1.8in}
%{\centering{\bf {P Deepa Shenoy}} is currently the Professor, University Visvesvaraya College of Engineering, Bangalore University, Bangalore. She obtained her Bachelor of Engineering from University Visvesvaraya College of Engineering. She received her Masters degree in Computer Science and} \\ \\
%\end{minipage}
%Automation from University Visvesvaraya College of Engineering, Bangalore University, Bangalore. She was awarded Ph.D. in Computer Science from Bangalore University, Bangalore.\\ \\

\noindent{\includegraphics[width=1in,height=1.8in,clip,keepaspectratio]{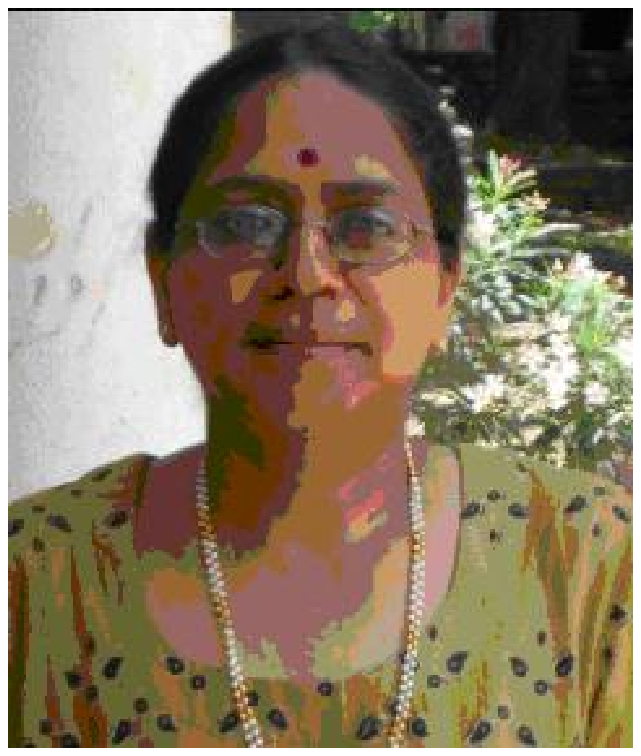}}
\begin{minipage}[b][1in][c]{1.7in}
{\centering{\bf {P Deepa Shenoy}} was born in India, on May 9, 1961. She graduated form UVCE, completed her M.E. from UVCE., has done her MS(Systems and information) from BITS., Pilani, and has obtained her Ph.D in CSE from Bangalore}\\
\end{minipage}
  University. She is presently employed as a Professor in department of CSE at UVCE. Her research interests include Computer Networks, Wireless Sensor Networks, Parallel and Distributed Systems, Digital Signal Processing and Data Mining. \\

\noindent{\includegraphics[width=1in,height=1.8in,clip,keepaspectratio]{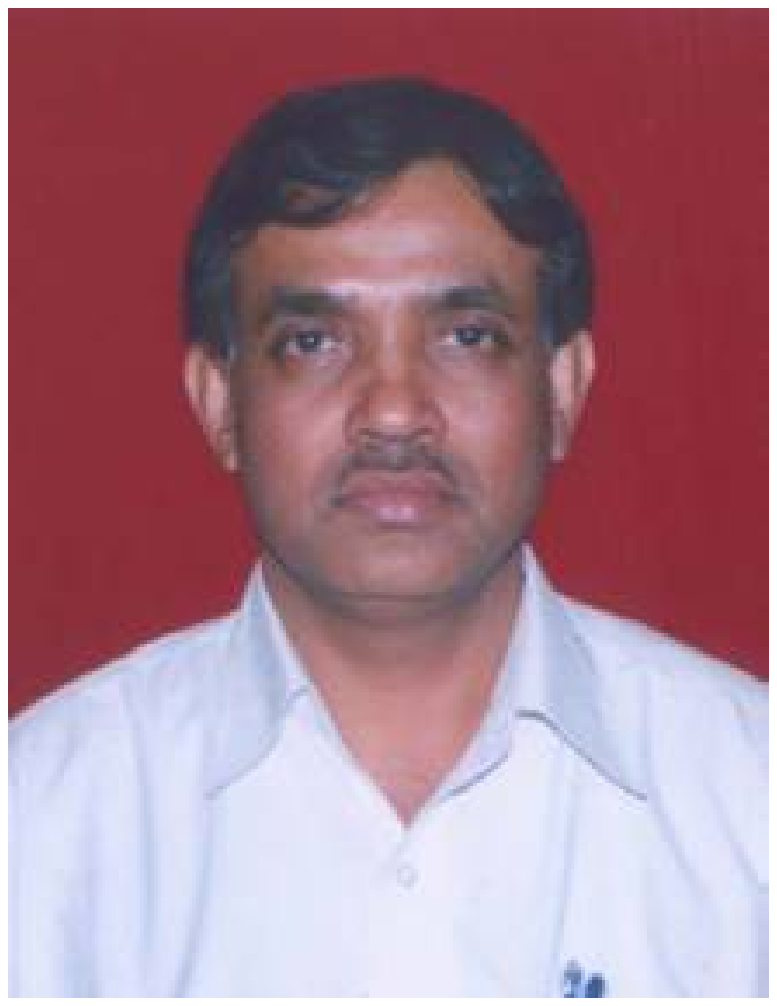}}
\begin{minipage}[b][1in][c]{1.8in}
{\centering{\bf {Venugopal K R}} is currently the Principal, University Visvesvaraya College of Engineering, Bangalore University, Bangalore. He obtained his Bachelor of Engineering from University Visvesvaraya College of Engineering. He received his Masters degree in Computer Science and} \\ \\
\end{minipage}
Automation from Indian Institute of Science Bangalore. He was awarded Ph.D in Economics from Bangalore University and Ph.D in Computer Science from Indian Institute of Technology, Madras. He has a distinguished academic career and has degrees in Electronics, Economics, Law, Business Finance, Public Relations, Communications, Industrial Relations, Computer Science and Journalism. He has authored and edited 39 books on Computer Science and Economics, which include Petrodollar and the World Economy, C Aptitude, Mastering C, Microprocessor Programming, Mastering C++ and Digital Circuits and Systems $etc.$. During his three decades of service at UVCE he has over 350 research papers to his credit. His research interests include Computer Networks, Wireless Sensor Networks, Parallel and Distributed Systems, Digital Signal Processing and Data Mining.\\ \\

\noindent{\includegraphics[width=1in,height=1.8in,clip,keepaspectratio]{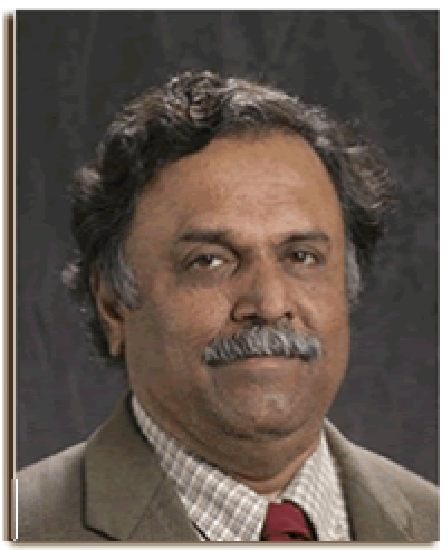}}
\begin{minipage}[b][1in][c]{1.8in}
{\centering{\bf {S S Iyengar}} is currently the Roy Paul Daniels Professor and Chairman of the Computer Science Department at Louisiana State University. He heads the Wireless Sensor Networks Laboratory and the Robotics Research Laboratory at LSU.  } \\ 
\end{minipage}
He has been involved with research in High Performance Algorithms, Data Structures, Sensor Fusion and Intelligent Systems, since receiving his Ph.D degree in 1974 from MSU, USA. He is Fellow of IEEE and ACM. He has directed over 40 Ph.D students and 100 Post Graduate students, many of whom are faculty at Major Universities worldwide or Scientists or Engineers at National Labs/Industries around the world. He has published more than 500 research papers and has authored/co-authored 6 books and edited 7 books. His books are published by John Wiley \& Sons, CRC Press, Prentice Hall, Springer Verlag, IEEE Computer Society Press $etc.$. One of his books titled Introduction to Parallel Algorithms has been translated to Chinese.\\ \\

\noindent{\includegraphics[width=1in,height=1.8in,clip,keepaspectratio]{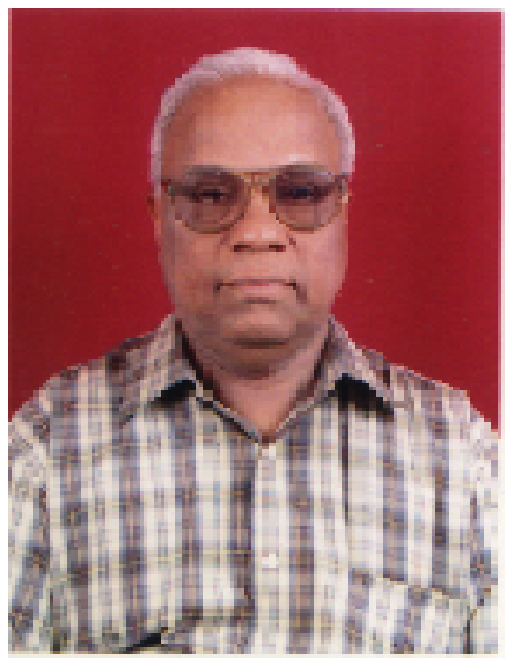}}
\begin{minipage}[b][1in][c]{1.8in}
{\centering{\bf {L M Patnaik}} is currently Honorary Professor, Indian Institute of Science, Bangalore, India. He was a Vice Chancellor, Defense Institute of Advanced Technology, Pune, India and was a Professor since 1986 with the Department of Computer Science and Automation, Indian} \\ \\
\end{minipage}
Institute of Science, Bangalore. During the past 35 years of his service at the Institute he has over 700 research publications in refereed International Journals and refereed International Conference Proceedings. He is a Fellow of all the four leading Science and Engineering Academies in India;  Fellow of the IEEE and the Academy of Science for the Developing World. He has received twenty national and international awards; notable among them is the IEEE Technical Achievement Award for his significant  contributions to High Performance Computing and Soft Computing. His areas of research interest have been Parallel and Distributed Computing, Mobile Computing, CAD for VLSI circuits, Soft Computing and Computational Neuroscience.
\balance

\end{document}